\documentstyle[twocolumn,amssymb,aps,psfig,floats]{revtex}

\begin{document}
\draft

\twocolumn[\hsize\textwidth\columnwidth\hsize\csname @twocolumnfalse\endcsname

\title{Insulator-Metal Transition in One Dimension \\
Induced by Long-Range Electronic Interactions}
\author{D. Poilblanc$^{1}$, S. Yunoki$^2$, S. Maekawa$^3$, and  E. Dagotto$^2$
}
\address{
$^{1}$Laboratoire de Physique Quantique \&
Unit\'e Mixte de Recherche 5626, C.N.R.S., \\
Universit\'e Paul Sabatier, 31062 Toulouse, France  \\
$^2$ Department of Physics and NHMFL, Florida State University,
Tallahassee, FL 32306, USA \\
$^3$ Institute for Materials Research, Tohoku University, Sendai 980-77, Japan
}

\date{April 97}
\maketitle

\begin{abstract}
\begin{center}
\parbox{14cm}{
The effects of a long range electronic 
potential on a one dimensional commensurate
Charge Density Wave (CDW) state are investigated. Using numerical
techniques it is shown that a transition to a metallic ground state
is reached as the range of the electron-electron repulsion increases.
In this metallic state, the
optical conductivity exhibits a large Drude weight. 
Possible interpretations of our results are discussed.
}
\end{center}
\end{abstract}

\vskip2pc]

As the dimensionality of an  electronic system decreases,
 it is expected that charge screening would become less important in
effectively
reducing the range of the electron-electron interactions. Thus, in one
dimension (1D)  the long-range (LR) Coulomb potential should play
an important role in determining the physical properties
of electronic models.\cite{kondo}
Experiments in
$GaAs$ quantum wires\cite{goni} and quasi-one-dimensional
 conductors\cite{dardel} 
indeed highlight the importance of LR 
interactions between carriers. Recently,
the role of a $1/r$ Coulomb repulsion on the long distance properties 
of electrons confined to a chain
has been theoretically investigated using 
bosonisation techniques~\cite{Schulz_crystal}. 
The 4$k_F$ charge correlations  decay very slowly with distance
suggesting that the ground state (GS) is 
similar to a classical Wigner crystal. 
However, this previous study was made in the continuum limit 
and the role of the lattice in this context is unclear.
In addition, theoretical\cite{giamarchi} 
and experimental\cite{optical} results suggest that
Umklapp processes have a crucial importance 
in 1D close to the Mott transition.

In order to investigate the interplay between the 
Coulomb interaction and Umklapp scattering on the phase diagram and transport 
properties of 1D systems, here static and dynamical
observables are calculated
on  chains with a variety of electron-electron interactions
using numerical techniques. 
As the range of the electronic potential increases,
an insulator to metal transition occurs.
In particular, 
for unscreened $1/r$ Coulomb
interactions on a lattice our computational analysis suggests that the
GS is {\it metallic}, an interesting result considering
that for short-range interactions (SR) and the densities studied here the
GS is a CDW insulator. 

The model analyzed here consists of a single chain with $L$-sites 
and an extended Hubbard-like interaction,
\begin{eqnarray}
H=-t\sum_{i,\sigma} (c_{i,\sigma}^\dagger c_{i+1,\sigma} + h.c.)~~~~~~~
\nonumber \\
+ \sum_{i\ge j} V_{i-j} (n_i-{\bar n})(n_j-{\bar n}),
\label{ham}
\end{eqnarray}
where $n_i$ ($\bar n$) is the local (average) electron density, and the
rest of the notation is standard ($t=1$ is the unit of energy).
$V_0=U/2$ corresponds to the on-site interaction ($U$ is the usual Hubbard 
local coupling) and will alway be included here. 
For $|i-j|\ge 1$, three types of potentials will be mainly considered:
(i) a plain LR Coulomb interaction
$V_{i-j}=V_C/|i-j|$;
(ii) a SR potential, typically $V_{i-j}=V_C/|i-j|$ for 
$|i-j|\le r_{max}$ 
and $V_{i-j}=0$ otherwise, extending up to a distance $r_{max}$ 
(generally with  $r_{max}=r_0-1$ where $r_0$ is the average 
inter-particle distance, i.e. $r_0=1/{\bar n}$ 
in units of the lattice spacing); and
(iii) a ``mixed'' potential with a Coulomb tail of adjustable intensity,
$V_{i-j}=V_C/|i-j|$ for $|i-j|\le r_{max}$ and $V_{i-j}=\beta V_C/|i-j|$ for 
$|i-j|> r_{max}$, $0<\beta<1$. Using Exact Diagonalization techniques,
1D chains are here analyzed.
A uniform ionic background has been included 
in Eq.~(\ref{ham}) such that the electrostatic energy per unit volume 
remains finite. 

\begin{figure}[hbt]
\begin{center}
\mbox{\psfig{figure=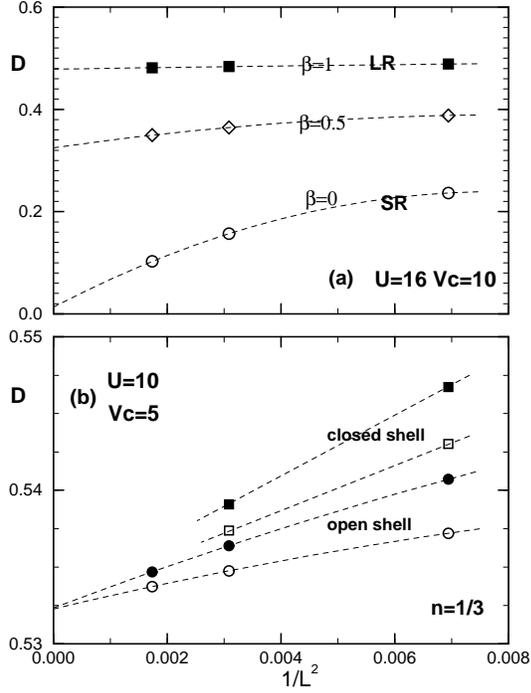,width=7cm,angle=0}}
\end{center}
\caption{
Charge stiffness $D$ vs $1/L^2$ for ${\bar n}=1/3$. Dashed lines correspond to 
polynomial fits of the form $a+b\frac{1}{L^2}+c\frac{1}{L^4}$.  
(a) Comparison between SR and LR potentials 
for open shell configurations. 
The SR potential extends up to $r_{max} = 2$;
(b) Comparison between $V_C^{\text sin}(r)$ (filled symbols) and 
$1/r$ (open symbols) potentials for  boundary conditions giving rise to
open or closed shell configurations. 
}
\label{drude}
\end{figure}

Let us  start
with a purely 1D commensurate CDW insulator
(reference state)
on which the LR Coulomb tail is switched on.
For commensurate densities and with SR
interactions, it is well-known that 
Umklapp processes can lead to an instability of the
Luttinger Liquid GS towards an insulating 
state.~\cite{Schulz_review1D} Let us explore model Eq.(1) with a SR potential 
using $r_{max}=r_0-1=2$ at density ${\bar n}=1/3$ varying 
the couplings $U$ and $V_C$ in order to determine a set of parameters 
for which the insulating nature of the GS can be unambiguously proven. 
Such insulating phase can be characterized by a vanishing charge stiffness:
if the electrons experience
a flux $\Phi$ (in unit of the flux quantum) threaded through the ring, 
the GS energy $E_0$ will become independent of $\Phi$ once the 
chain length exceeds some characteristic localization length. 
In Fig.~\ref{drude}(a) the charge stiffness 
$D=\partial^2(E_0/L)/\partial\kappa^2$ (where
$\kappa=\frac{2\pi}{L}\Phi$) is shown as a function of $1/L^2$ (a scaling
behavior also followed by Luttinger Liquid (LL) chains~\cite{LL_scaling}). 
In the case of the SR potential 
the data strongly suggests an insulating state since $D\rightarrow 0$ 
when $L\rightarrow\infty$ for the couplings chosen here.

\begin{figure}[htb]
\begin{center}
\mbox{\psfig{figure=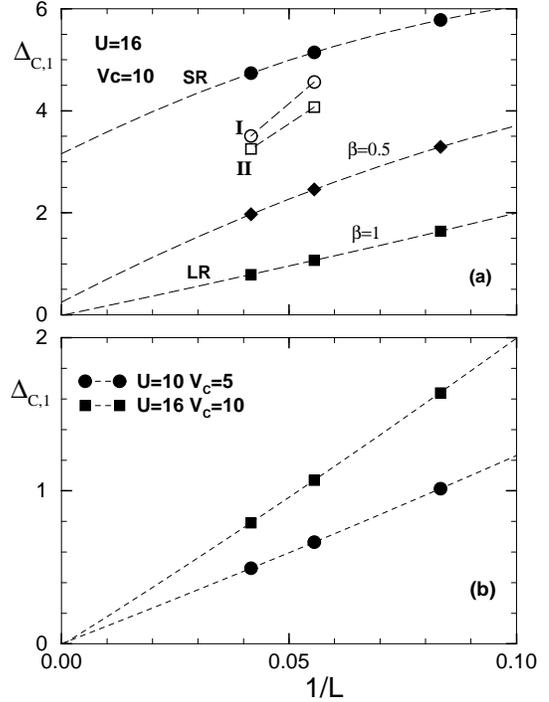,width=7cm,angle=0}}
\end{center}
\caption{
Finite size scaling of the single particle gap $\Delta_{C,1}$ 
using $U=16$ and $V_C=10$. Dashed lines correspond to 
polynomial fits of the
form $a+b\frac{1}{L}+c\frac{1}{L^2}$. 
(a) Comparison between SR and Coulomb (LR) interactions.
(b) $1/r$ Coulomb potentials. The positions of the 
absorption peaks labeled by I and II in Fig.~4(b) 
are also shown (open symbols).}
\label{SP_gap}
\end{figure}

The insulating character of this state can be further established from the
finite size scaling behavior of the single particle excitation 
energy (Fig.\ref{SP_gap}), defined as
$\Delta_{C,1}(L)=E_0({\bar n}L+1,L)+E_0({\bar n}L-1,L)-2E_0({\bar n}L,L)$, 
where $E_0(N_e,L)$ is the GS energy for $N_e$ 
electrons and $L$ sites~\cite{note_BC}.
For the SR potential and the same parameters 
as previously, Fig.~\ref{SP_gap}(a) strongly suggests that $\Delta_{C,1}$ 
extrapolates to a finite and sizable gap.
In addition, explicitly evaluating real space charge correlations in this state
a well-pronounced 
$4k_F=2\pi {\bar n}$ pattern was observed. The physical nature of this state 
is then clear: it forms a commensurate CDW state in which the particles
are localized on every $r_0$ sites. In the thermodynamic limit,
the GS is $r_0$-fold degenerate with gapped charge excitations\cite{comm1}.

\begin{figure}[hbt]
\begin{center}
\mbox{\psfig{figure=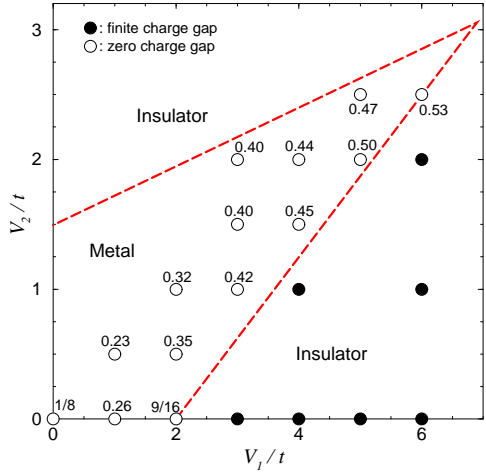,width=7cm,angle=-90}}
\end{center}
\caption{
Phase diagram of the $V_1$-$V_2$ model ($U=\infty$) at a density of 
${\bar n}=1/2$. See also Ref.~\protect\onlinecite{emery}.
The extrapolated values of the exponent $\alpha$ are indicated on the plot.
}
\label{Phase_diagram}
\end{figure}

Let us now consider a more extended interaction. The $U=\infty$
phase diagram of the repulsive $V_1$--$V_2$ model at quarter filling 
($\bar n=1/2$) obtained from the investigation of the charge gap $\Delta_{C,1}$
is shown in Fig.~\ref{Phase_diagram} ($V_1$ and $V_2$ are
the electronic density-density repulsion at distance 1 and 2,
respectively, while at larger distances the interaction is zero). 
For such a density ($r_0=2$),
the line $V_2=0$ corresponds to 
an insulating CDW state 
above a critical value of $V_1$ 
in agreement with our previous results for ${\bar n}=1/3$, and
with earlier calculations using spinless fermions 
by Emery and Noguera\cite{emery}.
When a finite $V_2$ is slowly switched on (now $r_{max}=r_0$)
the insulating phase is destroyed above a critical 
line. Thus, a repulsion beyond  
distance $r_0-1$ drives the system metallic
starting from a CDW insulator. Of special importance is
the case $V_2=V_1/2$,
which fulfills the relation between the first terms of the Coulomb
interaction, since it corresponds to a particularly stable metallic GS.
It is also important to stress that the metallic state is here of the
Luttinger Liquid (LL) type with $\omega_\rho (q) \sim u_\rho q$ charge 
excitations. The corresponding values of the LL 
exponent $\alpha$ characterizing the density of states~\cite{dardel}
($N(\omega)\sim\omega^\alpha$) are shown on Fig.3. This exponent
was obtained using $\alpha = (K_\rho + K^{-1}_\rho -2)/4$, and
$K_\rho$ was calculated from the compressibility and the charge
velocity following a standard procedure.\cite{schulz5}

Let us now turn to the case of a $1/r$ Coulomb potential. 
To facilitate the comparison with the previous results for the CDW
insulating regime,
 the density ${\bar n}=1/3$ at
$U=16$ and $V_C=10$ will be considered.
When the LR tail is added, in contrast to the SR potential, 
the scaling of the charge stiffness in Fig.~\ref{drude}(a) suggests
that it reaches a finite value as $L \rightarrow \infty$.
To check the accuracy of the finite size extrapolation
the sine approximation to the Coulomb potential
$V_C^{\text sin}(r)=\frac{V_C}{\pi L}/\sin{(r/\pi L)}$ was also
considered. 
A comparison between $V_C^{\text sin}$ and the plain $1/r$ interaction
is shown in Fig.~\ref{drude}(b).
All data sets are consistent with the same bulk
extrapolation. Then, the present results show
that the commensurate CDW state undergoes 
a transition to a metallic state as the range of the electronic interaction 
grows.

The metallic character of the $1/r$ Coulomb model can also be further
established from the finite size scaling of the single particle 
gap $\Delta_{C,1}$ as shown in Figs.~\ref{SP_gap}(a,b). For the
sets of parameters considered here, a robust $1/L$ scaling behavior is found which 
suggests that the charge gap vanishes in the bulk,
consistent with the finite value of the charge stiffness found before.
It is interesting to observe that using
a weaker Coulomb tail of amplitude $\beta V_c$
($\beta <1$) also drives the system into a metallic state as shown for
$\beta=0.5$ in Figs.~\ref{drude}(a) and \ref{SP_gap}(a). This suggests
that
the range of the interaction, rather than its intensity, is the key element
in inducing the insulator-metal transition. Also note
that the model with LR interactions is metallic apparently at any
value of $V_C/t$, i.e. an infinitesimal amount of kinetic
energy is enough to destroy the insulating state. This is to be 
contrasted with SR models, such as the $t-V_1$ model, where an
insulator-metal transition occurs only for a large enough hopping
amplitude\cite{emery}.
In addition, the metallic state 
exhibits collective spin modes 
similar to those of a LL, i.e. with a dispersion 
$\omega_\sigma(q)\sim u_\sigma q$. 
Using finite size chains
$u_\sigma$ can be easily calculated on  
closed shell configurations for $L=12$, $18$
and $24$ sites, by means of the energy difference between the first triplet state 
with momentum $2\pi/L$ and the GS.
We have found that $u_\sigma$ follows very closely
 a  $1/L^2$ scaling behavior, which
enables an accurate determination of its bulk value (table I).   
The small values of $u_\sigma$ indicates that the effective exchange 
coupling between neighboring particles  at a distance $r_0$ is
small.

\begin{figure}[htb]
\begin{center}
\mbox{\psfig{figure=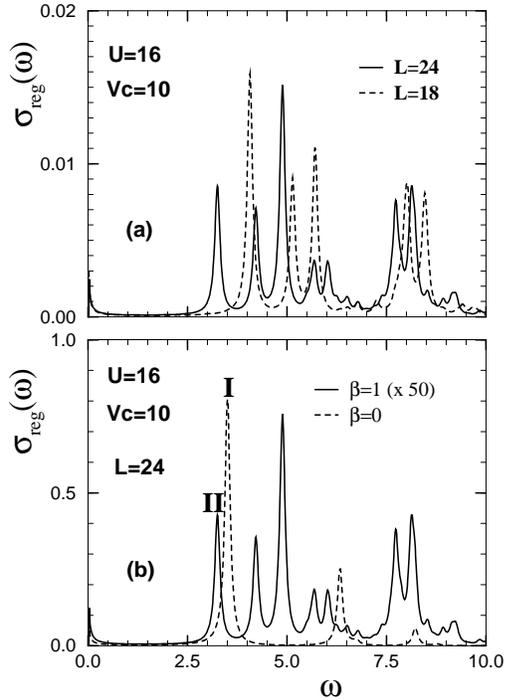,width=7cm,angle=0}}
\end{center}
\caption{
Optical absorption of a $1/r$-Hubbard chain. 
(a) Data for 18- and 24-sites chains.
(b) Comparison with the insulating CDW state ($\beta=0$).
The small absorption in the case of the 
LR Coulomb potential ($\beta=1$) has been multiplied by a factor 50
and the comparatively (very) large Drude peak at $\omega=0$ has been
omitted for clarity. 
}
\label{absorption}
\end{figure}

To gain insights on the transport properties of the metallic state 
observed here, the zero-temperature optical 
conductivity 
\begin{equation}
\sigma(\omega,T=0)=\pi D\ \delta(\omega) +\sigma_{reg} (\omega),
\label{cond}
\end{equation}
was numerically investigated.
Results for the LR potential are shown in Fig.~\ref{absorption}(a) for
$L=18$ and $L=24$ rings. The shape, as well as the magnitude of the 
absorption curve, depends weakly on the system size. By calculating 
separately $\sigma_{reg} (\omega)$, and using the previous results
for the charge stiffness $D$, it can be explicitly checked the
validity of the 
optical sum rule $\pi D/2 + I_0 = \frac{\pi}{2} E^0_K/L$, where $E^0_K/L$
is the kinetic energy per site and $I_0$ is the finite frequency 
integrated absorption. Extrapolations ($L\rightarrow\infty$) 
of these quantities are provided in Table I. 
It is clear that the metallic and insulating regimes
 have very different transport properties. Specially,
the weight of the absorption band of the metallic state (given
by $\sigma_{reg}(\omega)$) corresponds only to $1$ to $2\%$ 
of the total weight, while it exhausts the total sum rule in the case of the 
insulator\cite{giama}.
On the other hand, the energy scales of the absorption bands are comparable 
for the insulator and the metal (Fig.~\ref{absorption}(b)). 
With increasing system size the largest low energy peak in
$\sigma_{reg} (\omega)$ shifts to lower
energy (see a comparison with $\Delta_{C,1}$ in Fig.~\ref{SP_gap}) 
as its weight decreases. These excitations might correspond to
excitons which have lower energies than the CDW gap when finite range
interactions  are considered.

A simple interpretation of the insulator-metal transition 
observed here is as follows: consider the potential experienced
by a mobile electron on a chain assuming the rest of the electrons are fixed 
into a CDW configuration, with particles  equally spaced at
distances $r_0=1/{\bar n}$. For a strong SR interaction the
mobile electron is trapped in a deep square-well potential,
compatible with the insulating properties observed
numerically. However, as the potential range increases
the potential experienced by the mobile electron diminishes.
Actually, in the (unphysical) limit where the
repulsive potential between electrons is made
distance-independent
such potential becomes irrelevant.
This ``one-electron''
interpretation is favored by the fact that the insulator-metal
transition is not associated only with the (usually subtle)
$1/r$ interactions but it appears also for the SR $V_1-V_2$
model as well (Fig.~\ref{Phase_diagram}). 

\begin{table}
\begin{tabular}{|c|c|c|c|c|}
   & $u_{\sigma}$ & $\pi D/2$ & $I_0$ & sum \\
\hline
&&&&\\
$U=V_c=0$ & 1 & 1 & 0 & 1\\
$U=10$, $V_c=5$ $\beta=1$ &  0.13(5) & 0.836 & 0.010(3) & 0.846\\
$U=16$, $V_c=10$ $\beta=1$ & 0.04(1) & 0.752 & 0.017(1) & 0.769 \\
$U=16$, $V_c=10$ $\beta=0$ & 0.00(0) & 0.0 & 0.43 & 0.43 \\
&&&&\\
\end{tabular}
\caption{
Extrapolated values of the spin velocity $u_\sigma$, the Drude weight, 
the integrated optical absorption and the total sum rule 
for ${\bar n}=1/3$.
}
\end{table}

An alternative interpretation of our results
involves  a transition from a pinned CDW to a Wigner crystal
driven by the LR part of the potential\cite{note_4kF}.
The bosonisation calculations suggest that
the low frequency, long wavelength conductivity of the Wigner crystal
should behave as~\cite{Schulz_review1D},
\begin{equation}
\sigma(\omega,q) \propto 
\frac{i(\omega+i\epsilon)}{(\omega+i\epsilon)^2-(\omega_\rho(q))^2} \ .
\label{conductivity}
\end{equation}
In general, the $q\rightarrow 0$ and $\omega \rightarrow 0$ limits 
do not commute. The relations $\sigma(\omega,q=0)
=2D_{\text{Drude}}\delta(\omega)$ 
and $\sigma(0,q)=G\delta(q)$ define the Drude weight $D_{\text{Drude}}$
and the conductance $G$, respectively. 
Using $\omega_\rho(q)\propto q |\ln{q}|^{1/2}$ (plasmons),\cite{Schulz_crystal}
Eq.~(\ref{conductivity})
leads to a finite Drude weight and a vanishing conductance. 
The Drude weight is directly related to the 
charge stiffness calculated previously by $D_{\text{Drude}}=\pi D/2$. 
Therefore, our numerical results 
for the LR potential on the lattice are also
compatible with the 
bosonisation approach in the continuum limit, suggesting
that a metallic Wigner crystal GS can be realized 
on a lattice.~\cite{note} The metallicity is caused by 
collective excitations in this scenario. 

Quasi--1D organic conductors~\cite{Jerome_Schulz} like 
(TMTSF)$_2$PF$_6$ are compounds where $1/r$ Coulomb repulsion
might play an important role. This is suggested by 
ab-initio~\cite{Painelli} and Hartree Fock-Valence bond~\cite{Ducasse}
quantum chemical calculations.
Note also that despite its many successes,
 the description of quasi-one-dimensional conductors based on the LL approach
still faces various problems to provide a consistent explanation of {\it all} 
the experimental data 
such as NMR relaxation rates~\cite{NMR_Jerome}, 
photoemission~\cite{dardel} and IR absorption~\cite{optical}. 
As argued in this paper,  a long-distance 
$1/r$ Coulomb repulsion within the chains 
is expected to have drastic effects on the transport properties of 
such 1D systems, and this phenomenon may contribute to a more accurate
theoretical description of 1D conductors.

We thank IDRIS (Orsay) 
for allocation of CPU time on the C94 and C98 CRAY supercomputers.
Conversations with H. Schulz and V. Emery are acknowledged.
E. D. is supported by grant NSF-DMR-9520776. S. Y. acknowledges
financial support from the Japanese Society for the Promotion of Science.



\begin{references}
\bibitem{note_BC} 
Boundary conditions are such that ${\bar n}L$ non-interacting electrons 
on a $L$-sites chain form a closed shell configuration.
\bibitem{comm1}
However, low energy spin excitations still exist.
\bibitem{note_4kF} Explicit calculations show that the metallic state 
obtained here with the $1/r$ potential exhibits very strong
$4k_F$ charge correlations similar to those found in the insulating phase. 
However, in contrast to the CDW, we expect the charge-charge correlations 
to vanish at long distance in a WC.
\bibitem{note}
A weak pinning can not be completely excluded although 
it is not visible at the length scales considered here.
\end{references}
\end{document}